\documentstyle[aps,prl,preprint]{revtex}
\tightenlines
\input{psfig}

\newcommand{\beq}{\begin{equation}}
\newcommand{\eeq}{\end{equation}}
\newcommand{\bea}{\begin{eqnarray}}
\newcommand{\eea}{\end{eqnarray}}
\begin{document}
\draft
\title{Anisotropic and incommensurate spin fluctuations in hcp iron
and some other nearly magnetic metals.}
\author{V.Thakor\dag, J.B.Staunton\dag, J.Poulter\ddag, S.Ostanin\dag,
B.Ginatempo\P, and Ezio Bruno\P}
\address{\dag\ Department of Physics, University of Warwick, Coventry CV4~
7AL, U.K.}
\address{\ddag\ Department of Mathematics,Faculty of Science,
Mahidol University,Bangkok 10400,
Thailand.}
\address{\P\ Dipartimento di Fisica and Unita INFM, Universita di Messina, Italy}
\date{\today}
\maketitle
\begin{abstract}
We present an {\it ab initio} theoretical formalism for the static paramagnetic spin
 susceptibility of metals at finite temperatures. Since relativistic effects, 
e.g. spin-orbit coupling, are included we can identify the anisotropy or 
{\it easy axes} of the spin fluctuations. Our calculations
find hcp-iron to be unstable to in ab-plane, incommensurate anti-ferromagnetic (AFM) 
modes (linked to nested Fermi surface) below $T_N =$69K for the 
lowest pressures under which it is stable. $T_N$ swiftly drops to zero as the pressure is 
increased. The calculated susceptibility of yttrium is consistent with the helical, 
incommensurate AFM order found in many rare-earth-dilute yttrium alloys. Lastly, in line
with experimental data, we
find the easy axes of the incommensurate AFM and ferromagnetic spin fluctuations of 
the normal state of the triplet superconductor Sr$_2$RuO$_4$ to be perpendicular 
and parallel with the crystal c-axis respectively. 
\end{abstract}
At low temperatures under sufficient pressure iron  swaps its body-centered cubic
crystal structure in favor of a hexagonal close packed arrangement with the loss of its
ferromagnetic long range order. It also becomes superconducting~\cite{Shimizu} in a narrow range 
of  pressures between 15 to 30~GPa. This recent observation of superconductivity has added the 
phase of a common element to the growing list of magnetic or nearly magnetic materials exhibiting
a superconductivity in which magnetic fluctuations seem to play a key role~\cite{Mazin}. Prior 
to this finding the main interest in the magnetic properties of the high pressure hexagonal phase
of iron was motivated by their effect upon its structural and mechanical properties. As part of 
the study of the phase diagram of iron this has a wide spread relevance for subjects as disparate
as, for example, the properties of steel and the structure of the Earth's inner core~\cite{Gilder}.

The importance of magnetism upon the phase diagram is well-known. Ab-initio density
functional theory (DFT) calculations have demonstrated that the formation of magnetic
moments stabilises the b.c.c. phase of iron at low temperatures and normal pressures.
Although stable only above the Curie temperature, face-centered cubic iron
 has been found to have an incommensurate spin density wave (SDW) magnetic
state at low temperature following measurements on fcc-Fe precipitates on copper 
\cite{Tsunoda} and ab-initio calculations \cite{Uhl} which also show the magnetic interactions 
to vary sharply as a function of volume.

The magnetic attributes of hexagonal iron have not been so well identified. To date DFT 
calculations of the energies of two possible anti-ferromagnetic (AFM) ground states are lower than
the non-magnetic one for pressures up to 60 GPa and calculated elastic constants for such magnetic
states provide a better interpretation of experimental data than those from non-magnetic 
states~\cite{Steinle}. On the other hand, in-situ Mossbauer data from hcp iron \cite{Gort}
show no evidence of long range magnetism although they do allow for it to possess strongly 
enhanced paramagnetism which diminishes with increasing pressure. In this letter we study the 
spin fluctuations (SF) in this material via calculations of the paramagnetic spin susceptibility
as a function of both temperature and pressure. We find the predominant modes to be incommensurate
with the hexagonal lattice and set by nesting features of the Fermi surface. 
We find their 'easy axis' to be in the ab-plane. These fluctuations
become unstable at low temperatures and at the lower pressures 
where hcp-Fe is just stable, indicating an incommensurate SDW 
magnetic ground state. 

To carry out this study we describe a 
theoretical formalism to calculate a relativistic paramagnetic static spin 
susceptibility for metals at finite temperature.  We highlight how the
natural inclusion of 
spin-orbit coupling allows us to study the anisotropy of enhanced 
spin fluctuations in nearly magnetic systems and to identify 
their $\emph{easy-axes}$. 
The paramagnetic susceptibility  of a metal can possess much structure in wave-vector (${\bf q}$) space.
Indeed
a peak at one temperature can evolve into a divergence at a lower one to signify
the system's transition into a magnetically ordered state characterised by a static
magnetisation wave with a wave-vector corresponding to that of the peak. By seeking out the
peaks of the paramagnetic susceptibility by examining a wide range of wave-vectors we
can identify the dominant spin fluctuations and probe for many different possible magnetic 
ground states.
 Moreover the
easy axis of the enhanced spin fluctuations in the 
paramagnetic phase will indicate the direction along which magnetisation grows in the 
ordered phase.

Following the application to hcp-Fe, we calculate the ${\bf  q}$-dependence and
easy axis of spin-fluctuations in hcp metals yttrium and scandium.
Our results are consistent 
with the helical anti-ferromagnetic structure that occurs when Y is doped with magnetic rare earth impurities such as Gd. 
Finally we investigate the anisotropy of the spin fluctuations in the normal state
of the perovskite spin-triplet superconductor Sr$_2$RuO$_4$. This aspect has
been proposed as an ingredient for the theoretical understanding of the pairing mechanism in
this extensively studied material \cite{Kuwabara}. Our principal findings are
 in good agreement with NMR data~\cite{Mukuda}.

We access the paramagnetic spin susceptibility by considering a paramagnetic metal 
subjected to a small, external, inhomogeneous magnetic field,
$\delta {\bf b}^{ext.}({\bf r})$, which induces a magnetisation
 $\delta {\bf m} ({\bf r})$. We use DFT to derive an expression via a variational linear
 response approach~\cite{Vosko+Perdew,Applied_to_TM}. Although there are a 
number of studies of this type~\cite{SW+Sav}, relativistic effects are typically
 omitted. We address this issue in this letter. 
Recently we have developed~\cite{staun1} a scheme for calculating the wave-vector,  
frequency and temperature-dependent dynamic spin susceptibility and have   
applied it to Pd, Cr and Cr-alloys obtaining good agreement 
with experimental data. Here we describe the new aspects that emerge when
relativistic effects are incorporated in its static limit. 

We start with the relativistic version of density functional theory~\cite{Rajg+Call} 
where a Gordon decomposition is applied to the current density and the
spin-only part of the current retained. 
The Kohn-Sham-Dirac equation involving effective one-electron fields for a paramagnetic
 system is solved by a one-electron Green's function $G_o({\bf r} , {\bf r}' ; \varepsilon)$.  
If a small external field $\delta {\bf b}^{ext}$ is applied 
along a direction $\hat{{\bf n}}$ with respect to the crystal axes,
a small magnetisation, $\delta {\bf m} ({\bf r})$, and 
effective magnetic field are set up. The latter is
$\delta {\bf b}^{eff}[\rho({\bf r}),{\bf m}({\bf r})] = 
\delta{\bf b}^{ext}({\bf r}) + I_{xc}({\bf r})\delta {\bf m}({\bf r})$   
where we have assumed that  $\delta {\bf b}^{ext}$ couples  only to the spin part 
of the current and $I_{xc}({\bf r})$ is the functional derivative of the effective exchange and correlation 
magnetic field (within the local density approximation e.g.~\cite{vBH} LDA) with 
respect to the induced magnetisation density.
The response of the system to the external magnetic field is expressed in terms of the
Green's function obtained from a Dyson equation in terms of unperturbed Green's function 
$G_o({\bf r} , {\bf r}' ; \varepsilon)$ of the paramagnetic system and  
$\delta{\bf b}^{eff}$.
For a general crystal lattice with $N_s$ atoms
located at positions                                               
${\bf a}_l$ ( $l$ = 1,..,$N_s$) in each unit cell,
a lattice Fourier transform can be carried out and
an expression for the full static spin susceptibility found.
\begin{eqnarray} 
\chi^{{\hat{\bf n}}} ({\bf x}_{l},{\bf x}'_{l'}\, ,{\bf q}) \, = \,
\chi_{o}^{{\hat{\bf n}}} ({\bf x}_{l},{\bf x}'_{l'}\, ,{\bf q}) + 
%\qquad \qquad \qquad \qquad \qquad \qquad \nonumber \\
\sum_{l''}^{N_{s}} \int \, 
\chi_{o}^{{\hat{\bf n}}} ({\bf x}_l,{\bf x}''_{l''}\, ,{\bf q})
\, I_{xc}({\bf x}''_{l''}) \,
\chi^{{\hat{\bf n}}} ({\bf x}''_{l''},{\bf x}'_{l'}\, ,{\bf q})
\, d{\bf x}''_{l''}  
\label{eq:X}
\end{eqnarray}    
with the non-interacting susceptibility of the static unperturbed system given by
\begin{eqnarray} 
\chi_o^{{\hat{\bf n}}} ({\bf x}_l,{\bf x}'_{l'}\, ,{\bf q}) \, = 
-(k_B T) \,Tr \, \tilde{\beta}\, \mbox{\boldmath${\tilde{\sigma}}$}.{\hat {\bf n}} \,
\sum_n 
%\qquad \qquad  \qquad \qquad \qquad \qquad \qquad \nonumber \\ 
\int \, \frac{d{\bf k}}{\nu_{BZ}}  
G_o({\bf x}_l,{\bf x}'_{l'}\,,{\bf k}, \mu+i\omega_n)\,
\tilde{\beta}\, \mbox{\boldmath${\tilde{\sigma}}$}.{\hat {\bf n}} \,
G_o({\bf x}'_{l'},{\bf x}_{l}\,,{\bf k}+{\bf q}, \mu+i\omega_{n})  
\label{eq:Xo}
\end{eqnarray} 
The integral is over the Brillouin zone with wave vectors ${\bf k}$, ${\bf q}$ and 
${\bf k}+{\bf q}$ within the Brillouin zone of volume $\nu_{BZ}$. 
The ${\bf x}_l$ are measured relative to the positions of atoms centred on ${\bf a}_l$. 
The fermionic Matsubara frequencies $\omega_{n}$ are ${(2n+1) \pi k_{B}T}$ and 
$\mu$ is the chemical potential. The Green's function for the unperturbed,
 paramagnetic system containing the band structure effects
is obtained via relativistic multiple scattering (Korringa-Kohn-Rostoker, KKR)
theory~\cite{Faulkner+Stocks}. The susceptibility has a dependence on the direction
 of the  magnetic field, ${\hat{\bf n}}$, which is lost when relativistic, spin-orbit
coupling effects are omitted. If an external magnetic field is applied
along the $(0,0,1)$ direction, i.e. the c-axis in h.c.p. and tetragonal systems,  
 we have $\chi^{z}$ whereas $\chi^{x}$ is given for the field in the ab-plane,
${\hat{\bf n}}=(1,0,0)$. 
We obtain an anisotropy as the difference in the susceptibility
 when an external magnetic field is applied in two distinct directions with respect to 
the crystal axes i.e. ($\chi^{x} - \chi^{z}$).
 This approach presented here is applicable to ordered 
compounds and elemental metals and can be modified to study disordered alloys~\cite{staun1}.

Equation (\ref{eq:X}) is solved using a direct method of matrix inversion. The full Fourier
transform is then generated $\chi^{{\hat{\bf n}}} ({\bf q},{\bf q}) \, = \,
(1/V)\sum_{l} \, \sum_{l'}\, e^{i{\bf q}.({\bf a}_{l}-{\bf a}_{l'}')} \,
\int d {\bf x}_{l} 
\int d {\bf x}'_{l'} \,
e^{i{\bf q}.({\bf x}_{l}-{\bf x}_{l'}')} \,
\chi^{{\hat{\bf n}}} ({\bf x}_{l},{\bf x}'_{l'}\, ,{\bf q})$ where $V$
is the volume of the unit cell. Some aspects of 
the numerical methods used to evaluate (\ref{eq:Xo}) and (\ref{eq:X}) can be found 
in~\cite{staun1,Ben+Ezio} and further details are given elsewhere~\cite{Vijay}. 
We have used atomic sphere approximation (ASA), effective one-electron potentials
and charge densities in the calculations for hcp-Fe, Y and Sc. For Sr$_2$RuO$_4$ we have used 
muffin-tin (MT) ($l$=0) components of the potentials from a full potential calculation~\cite{Wien97}. 
In all cases the details of the electronic structures including the Fermi surfaces 
compare well with those from full potential calculations. 

Figure 1 shows some results of our susceptibility calculations for hcp-Fe
for four volumes in the range (128$a_{0}^3 \,< \,$V$\,<\,$143$a_{0}^3$) where
$a_{0}$ is the Bohr radius. The largest volume we consider is 142.23
$a_{0}^3$
which corresponds to a pressure of P$\sim$16GPa~\cite{Basset+Huang} and the smallest
 volume is 128.98 $a_{0}^3$ or  P$\sim$45GPa.
For all the volumes the axial ratio of the lattice constants is taken to be $c/a = 1.6$
 which is very close to the experimentally 
observed value over a wide pressure range~\cite{Basset+Huang}.
Calculations of $\chi({\bf q})$ were performed for many wave-vectors in the
 Brillouin zone in a thorough
search for dominant spin fluctuations and potential magnetically ordered states.
The important ones are shown in Fig.1. The temperature range for such calculations was 
50K$\,\leq\,$T$\,\leq\,$300K. 
According to the full potential DFT calculations of Steinle-Neumann 
{\it et al}~\cite{Steinle}, two AFM 
configurations (termed AFM-I and AFM-II) are more stable than non-magnetic
 or FM configurations for pressures up to $\sim$~60GPa at $T=0$K. AFM-I has
 magnetisation alternating in orientation in
ab-planes stacked along the c-axis while
AFM-II has magnetisation with opposite orientation on each layer 
aligned with the c-axis and perpendicular to the x-axis.
If AFM-I were the ground state magnetic configuration then for temperatures above 
the N$\acute{\textrm{e}}$el temperature, $T_N$ we should see a peak in the susceptibility  
at the special point wave-vector ${\bf q}_{A}= (0,0,a/2c)$, (in units of $2 \pi/a$).
 Similarly if AFM-II was the ground state configuration
then we expect to see the peak in $\chi({\bf q})$ at the special point ${\bf q}_{M}= 
( 1/\surd 3,0,0) $. 
For all four volumes, however, we find the maximal peak in $\chi({\bf q})$ to be at an
{\it incommensurate} wave-vector lying in the 
basal plane of ${\bf q}_{inc.}$=$(0.56,0.22,0)$. We can trace this
wave-vector ${\bf q}_{inc.}$ to a nesting of the metal's Fermi surface as shown in 
Fig.2. This is a cross-section of the Fermi surface which is generated from
 4 bands straddling the Fermi energy and is comprised of four sheets. 
A cross-section through the basal plane reveals the Fermi surface to be 
dominated by  two hexagonal-like shapes centred on $\Gamma$. 
The incommensurate ${\bf q}_{inc.}$=$(0.56,0.22,0)$ 
nests two pieces of Fermi surface as shown in Fig.2 which 
is a cross-section  where the x-axis is along the direction of ${\bf q}_{inc.}$ 
and the y-axis is along the c-axis.  

These results suggest that for the larger volumes the ground state of hcp-Fe is an 
incommensurate AFM. As the pressure is increased the N$\acute{\textrm{e}}$el temperature decreases 
to zero. Figure 1 shows that at 100K enhanced AFM-SF exist which decrease in
strength as volume decreases with increasing pressure. Further calculations of the full
 spin susceptibility, eq.(\ref{eq:X}), reveal that an instability 
arises at the nesting vector ${\bf q}_{inc.}$ above 50K at  $T_N=$ 69K
for the volume V = 142.23 ($a_{0}^3$) (P$\sim$16GPa) 
where hcp-Fe is stable.  $T_N$ drops below 50K (the lowest temperature we can consider) 
for the other smaller volumes.
 We therefore infer that hcp-Fe has
an incommensurate SDW ground state in a small pressure range starting at the onset of the hcp-phase
$\sim$15 GPa. Our calculations of the anisotropy reveal
$\chi^{x}_{o}({\bf q}_{inc.}) > \chi^{z}_{o}({\bf q}_{inc.})$
for all 4 volumes and all temperatures in the range (50K$\leq$T$\leq$300K). Thus
hcp-Fe exhibits incommensurate AFM-SF in the ab-plane.
If the superconductivity of hcp-Fe is unconventional 
as has been suggested~\cite{Mazin} then the in-plane incommensurate AFM-SF
must play an important role in the pairing mechanism.

Fig.1 also shows how $\chi^{-1}({\bf q})$ varies with volume for
other salient wave-vectors at T = 100K.
We see that $\chi({\bf q}_{inc.})>\chi({\bf q}_K)>\chi({\bf q}_M)>
\chi({\bf q}_A)>\chi({\bf q}_o \simeq (0,0,0))$ for all 4 volumes.
Apparently hcp-Fe is furthest away from forming a ferromagnetically 
ordered state as shown by the relatively high value of $\chi^{-1}({\bf q}_o)$.
This is followed by AFM-I (${\bf q}_{A}$) and then AFM-II (${\bf q}_{M}$). 
In agreement with~\cite{Steinle} we find that the AFM-II is a more stable
 configuration than AFM-I but we find a third special-point antiferromagnetic
 structure, AFM-III, characterised by ${\bf q}_K = (1/\surd 3, 1/3, 0)$, to be more
 stable than either of these. This structure
would have the magnetisation direction alternating in layers stacked along ${\bf q}_K$.   
Although ${\bf q}_{K}$ signifies a more stable AFM 
configuration than either AFM-I and AFM-II, it is at the incommensurate vector 
(${\bf q}_{inc.}$) where we see the maximal peak in $\chi({\bf q})$ which leads
 to our prediction that hcp-Fe has an incommensurate SDW ground state. 

Having established a way of determining the easy axis for spin fluctuations with our
relativistic formalism, we can use it to deduce how magnetic impurities are likely to
orientate in paramagnetic host metals. Excellent case studies are 
provided by yttrium alloys with low
concentrations of rare earths. When crystal field effects of the rare earths do not 
dominate, the enhanced susceptibility of Y determines the magnetic ordering. Here we 
describe our calculations of this for Y and also its lighter 3d counterpart Sc.
Both metals have hcp crystal 
structures with very similar electronic band structures and Fermi surfaces. 
We have used experimental lattice constants $a =6.89$, $c=10.83$ for Y and 
$a=6.24$, $c=9.91$ for Sc, in atomic units, $a_{0}$.
Our calculations of the static spin susceptibility defined by (\ref{eq:X}) show a peak at the 
wave-vector ${\bf q}_{inc.} = (0,0,0.18) 2\pi/a =(0,0,0.57\pi/c)$ for both Y and Sc.
 This vector ${\bf q}_{inc.}$ once again corresponds to a
 nesting effect from a webbing feature of flat parallel sheets~\cite{YExp_N} as 
confirmed by Fermi surface calculations and is in good agreement with 
experiments~\cite{YExp_N,YExp_N2}. 
Fig.3 shows the temperature dependence of the anisotropy of Y and Sc at the nesting vector 
${\bf q}_{inc.}$ and also at a small vector ${\bf q}_o\simeq (0,0,0)$
 (indicating ferromagnetic spin 
fluctuations (FM-SF)). Evidently the
anisotropy is an order of magnitude larger in Y that in Sc - e.g.  at T = 50 K 
the anisotropy 
$(\chi_o^{x}({\bf q}_{inc.})-\chi_o^{z}({\bf q}_{inc.}))$ for Y is 0.0411 ($\mu_{B}^2/Ry.$) 
while for Sc it is 0.00635 ($\mu_{B}^2/Ry.$).  
This difference is caused by the  spin-orbit
coupling being more pronounced in the heavier 4d metal Y than in the 3d Sc. 
At ${\bf q}_{inc.}$, the AFM-SF have an easy direction in the ab-plane, 
$\chi_{o}^{x}({\bf q}_{inc}) > \chi_{o}^{z}({\bf q}_{inc.})$, whereas
 that of the FM-SF is 
along the c-axis, $\chi_{o}^{z} > \chi_{o}^{x}$. The
anisotropy is also larger at ${\bf q}_{inc.}$ than for small wave-vectors. 
These calculations show that both Y and Sc exhibit AFM-SF characterised by the 
incommensurate wave-vector ${\bf q}_{inc} = (0,0,0.57 \pi/c)$ and are
aligned in the basal plane. They explain why when Y is perturbed by the addition of 
rare earth magnetic impurities (e.g. Gd), 
it typically responds by ordering the impurity moments into a
helical AFM state in which the moments align in the basal plane and rotate
their orientations in successive planes around the c-axis~\cite{Y-Gd}.  

We conclude with our study of the anisotropy of the spin-fluctuations of the normal 
state of Sr$_2$RuO$_4$. We have used lattice constants of 
$a=7.30$, $c=24.06$ in $a_0$ for the tetragonal unit cell.
Our calculations of the susceptibility and Fermi surface for Sr$_2$RuO$_4$ show an 
incommensurate nesting vector of ${\bf q}_{inc.}$=$(0.35,0.35,0)$ similar to 
that found in previous
calculations~\cite{Mazin2} and experiment~\cite{INS_data}. We have compared the 
anisotropy at the incommensurate wave-vector ${\bf q}_{inc}$ with that at small 
wave-vectors describing ferromagnetic SF. For the former we find
$(\chi_o^{z}({\bf q}_{inc.}) > \chi_o^{x}({\bf q}_{inc.}))$ implying 
that the incommensurate AFM-SF should have an 
easy direction along the $\emph{c-axis}$. Conversely for small ${\bf q}$ we find 
$(\chi_o^{x} > \chi_o^{z} )$ and therefore the FM-SF to be $\emph{in the ab-plane}$. These 
findings are in good agreement with NMR experiments~\cite{Mukuda}. At T = 50 K, the anisotropy
$(\chi_o^{z}({\bf q}_{inc.}) - \chi_o^{x}({\bf q}_{inc.}))$ is 0.847 ($\mu_{B}^2/Ry.$) which is an order of
magnitude greater than what we find for Y owing to the large tetragonal distortion (c/a = 3.3) 
of the unit cell. Finally a calculation of $(\chi_o^{x}({\bf q}_{inc.}) /
 \chi_o^{z}({\bf q}_{inc.}))$ at T=50K
yields  0.98. According to a scenario based on a simple model by Kuwabara
 $\it{et\,al}$~\cite{Kuwabara}, the spin-triplet superconducting state 
of Sr$_2$RuO$_4$ can be stabilised by incommensurate  AFM-SF 
with the easy axis for the fluctuations along the c-axis. Our value of the
anisotropy of the SF, $(\chi_o^{x}({\bf q}_{inc.}) / \chi_o^{z}({\bf q}_{inc.}))$, of 
 0.98, however, would also imply that the 
system must be very close to ordering magnetically.

We acknowledge support from the EPSRC (UK).

\newpage
\begin{figure}
\begin{center}
\psfig{figure=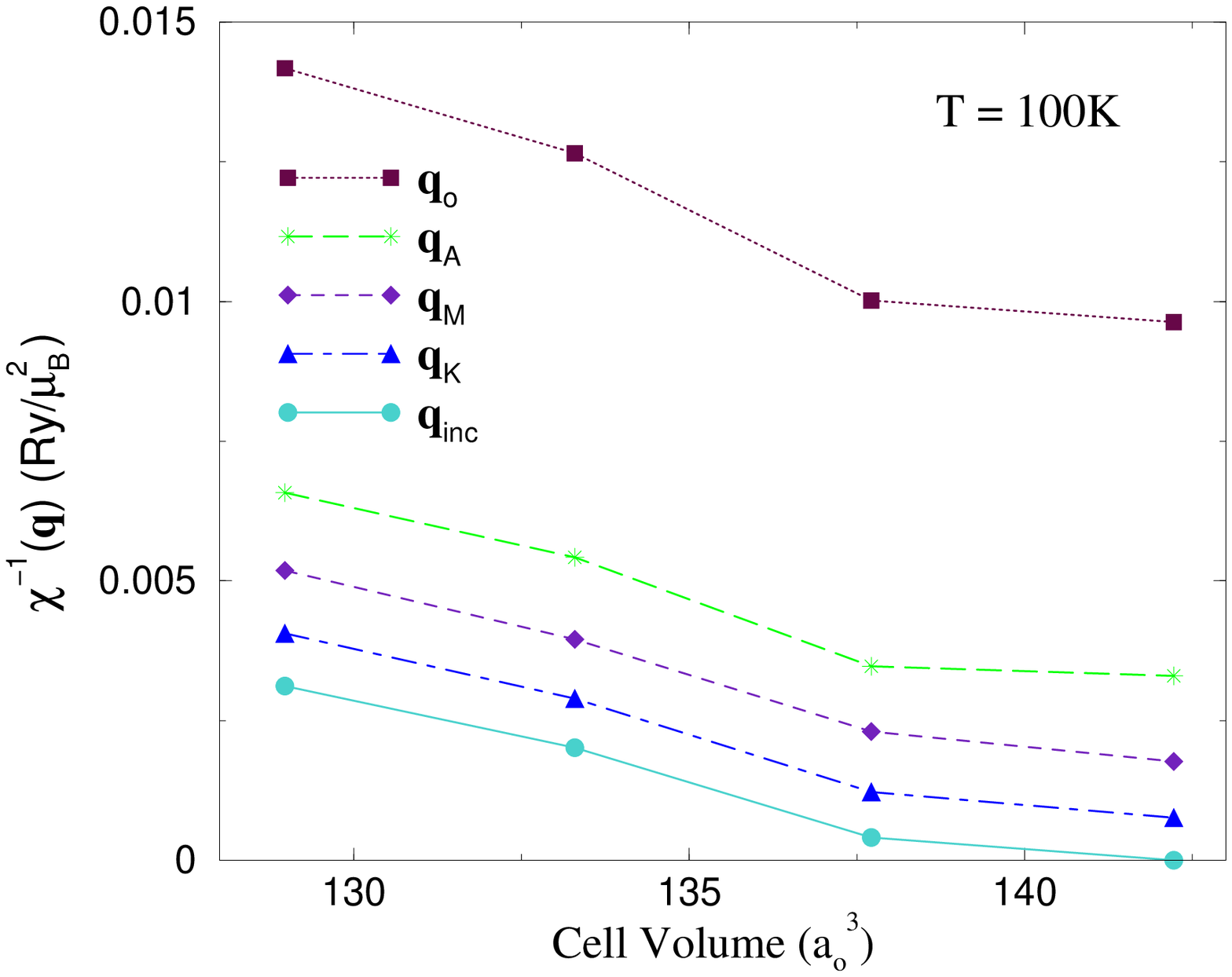}
\end{center}
\caption{The inverse susceptibility of hcp-Fe at $T=$ 100K for various wave-vectors 
(${\bf q}_{o}\simeq 0$, the nesting 
vector ${\bf q}_{inc.}=2\pi/a(0.56,0.22,0)$ and at the special points A, M and K) 
for a number of unit cell volumes.}
\end{figure}
\newpage
\begin{figure}
\begin{center}
\psfig{figure=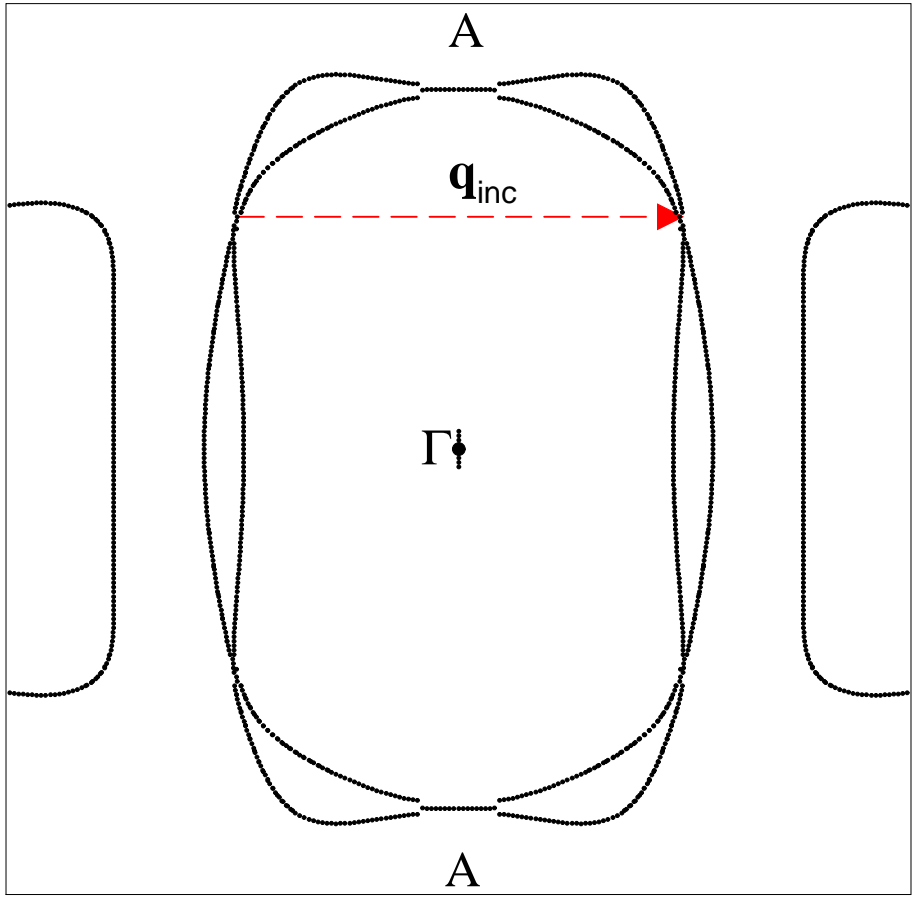}
\end{center}
\caption{Cross-section of the Fermi surface of hcp-Fe at V=137.72 $a_0^3$  
where the x- and y-axes are  along  ${\bf q}_{inc.}= 2\pi/a(0.56,0.22,0)$
 and the c-axis $(0,0,1)$ respectively.}
\end{figure}
\newpage
\begin{figure}
\begin{center}
\psfig{figure=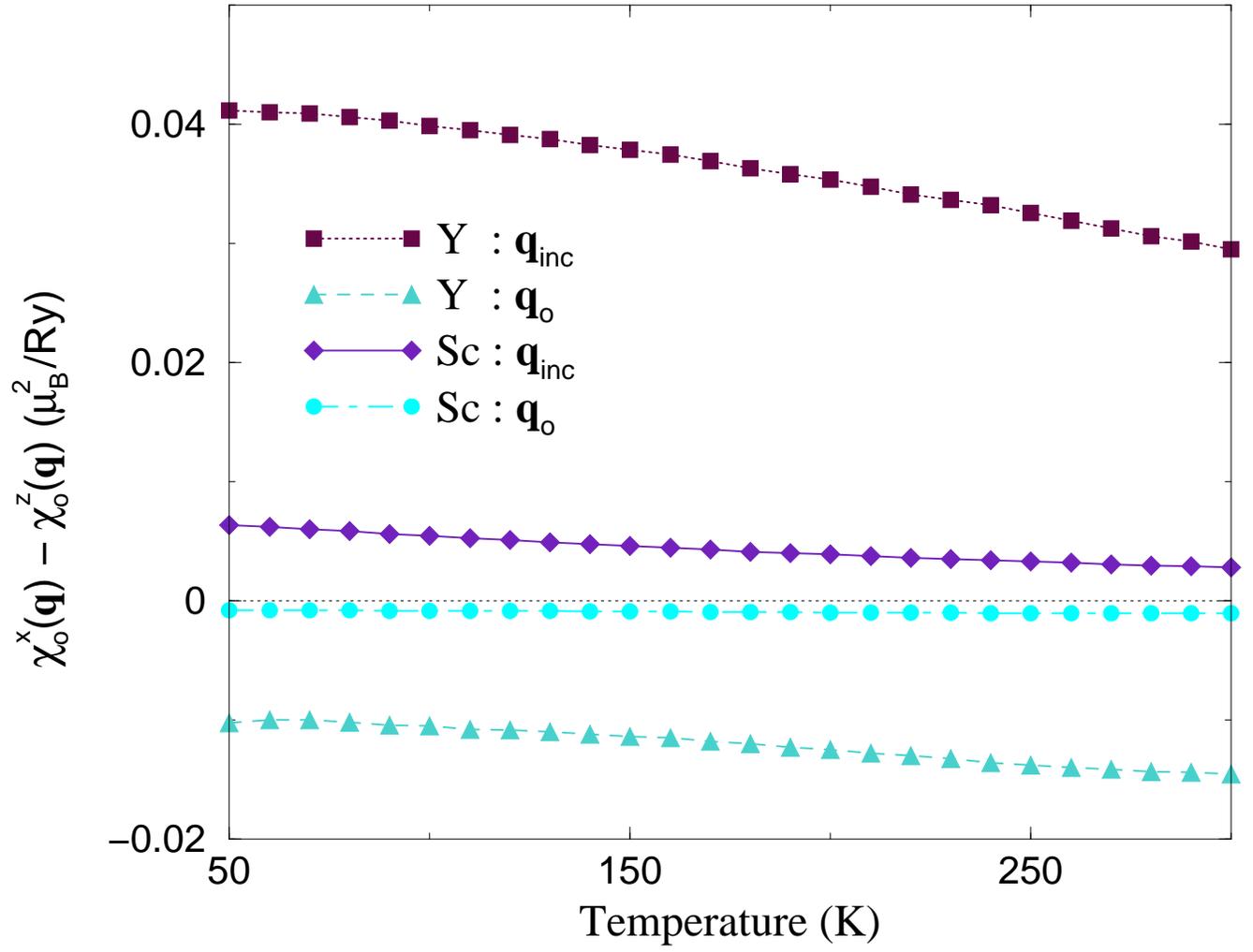}
\end{center}
\caption{The temperature dependence of the anisotropy $\chi_o^{x}({\bf q})
-\chi_o^{z}({\bf q})$
of Y and Sc at both the nesting vector ${\bf q}_{inc.}=(0,0, 0.57\pi/c)$ 
and ${\bf q}_{o}\simeq (0,0,0)$.}
\end{figure}
\end{document}